# Achieving Robust Extrapolation in Materials Property Prediction via Decoupled Transfer Learning

Tasuku Sugiura[1] and Teruyasu Mizoguchi[1]

[1]Institute of Industrial Science, The University of Tokyo, 153-8505 Tokyo, Japan

**Abstract**

Machine learning has revolutionized materials property prediction, yet fails catastrophically when extrapolating beyond training distributions—precisely the capability required for discovering unprecedented materials. Graph neural networks (GNNs) exhibit this collapse because end-to-end training fundamentally couples learned representations to target property distributions, preventing genuine extrapolation. We demonstrate that decoupled transfer learning—separating pretrained GNN feature extractors from simple regressors—overcomes this barrier. Pretrained features provide transferable structural knowledge, while simple regressors enable smooth extrapolation by maintaining learned trends beyond training boundaries. Benchmarked on layered intercalation compounds through four rigorous extrapolation scenarios and a temporal Materials Project split, our framework achieves 68% error reduction (RMSE: 0.881 vs. 2.778 eV/atom) versus end-to-end GNNs for extrapolation. Failure analysis reveals extrapolation succeeds for continuous chemical space but fails for discontinuous space, establishing clear design principles. Validated on Fermi energy prediction, this framework is immediately deployable using existing pretrained models, requiring no architectural innovations—transforming ML-driven materials discovery.

**Graphical Abstract**

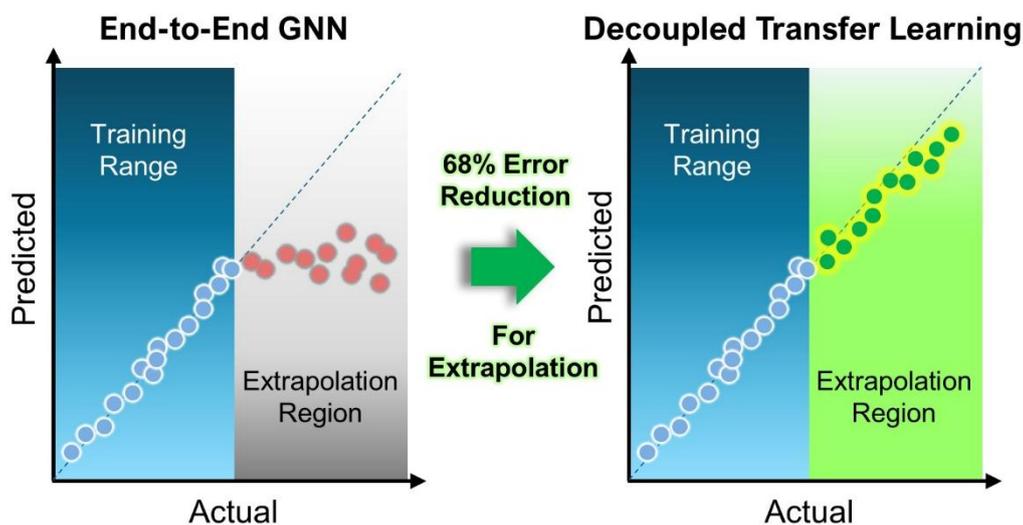

**Introduction**

Accelerating materials discovery is essential for addressing global challenges in energy storage, catalysis, and sustainable technologies[1–6]. Computational high-throughput screening has emerged as a powerful approach to identify promising candidates before costly experimental synthesis, but its success depends critically on accurate property prediction across vast chemical spaces. Recent advances in machine learning, particularly graph neural networks (GNNs) that represent crystal structures as atomic graphs, have achieved remarkable accuracy in predicting material properties[7–15]. The emergence of foundation models, pretrained on millions of structures from large-scale datasets such as the Materials Project[2] and Open Catalyst Project[16], has further enhanced these capabilities by creating versatile feature extractors applicable to diverse prediction tasks[9,17,18].

This progress conceals a fatal flaw: all current approaches collapse when predicting materials beyond their training distribution. This isn't a minor limitation—it's a fundamental barrier preventing ML from fulfilling its promise. Materials discovery fundamentally requires predicting properties of compounds that differ substantially from known materials, either exploring uncharted chemical spaces or targeting performance that exceeds existing compounds. Yet machine learning models, despite achieving high accuracy on random splits, exhibit catastrophic performance collapse under extrapolative conditions[19–23]. Studies employing realistic evaluation schemes, such as temporal splits where models predict properties of newly proposed materials or compositional splits reflecting discovery constraints, have revealed this critical weakness[20,21,24–26]. This extrapolation failure fundamentally undermines the central promise of machine learning for materials discovery: guiding experimental efforts toward genuinely novel, high-performance materials.

Various strategies have been proposed to address this challenge. Models using physically meaningful descriptors and simple approaches such as linear models can achieve effective extrapolation[17,22,24,27]. However, these methods typically rely on problem-specific feature engineering grounded in well-understood physical frameworks, limiting their applicability across diverse materials tasks. Meanwhile, complex neural networks achieve superior interpolation accuracy but fail catastrophically at extrapolation. This fundamental trade-off between generalizability, accuracy, and extrapolation capability has remained unresolved.

Here, we demonstrate that decoupled transfer learning—separating representation learning from property prediction—breaks this trade-off, achieving both high interpolative accuracy and robust extrapolation capability. The insight is counterintuitive: the very feature that enables deep learning's success—end-to-end optimization—is what prevents extrapolation. Joint optimization of feature extraction and property prediction creates representations that encode not only structural patterns but also implicit distributional constraints from target properties. This coupling locks model outputs within training ranges, creating an invisible barrier to extrapolation. Our decoupled approach breaks this constraint by transferring pretrained GNN representations, learned from millions of diverse structures, to simple regression models that naturally extend predictions beyond training ranges. This decoupling preserves the rich structural knowledge from foundation models while unlocking genuine extrapolation through the mathematical properties of linear-like regression.

We validate this approach through comprehensive evaluation across multiple extrapolation scenarios using layered intercalation compounds—a materials class critical for battery technologies[28,29]. We systematically assess performance under four conditions that mirror real discovery challenges: (1) interpolation (random splits—the standard but unrealistic baseline), (2) structural extrapolation (predicting materials novel host structures never seen during training), (3) property extrapolation (targeting extreme formation energies far outside training ranges), and (4) coupled extrapolation (the ultimate test: simultaneous structural and property extrapolation). Benchmarking against temporal alloy datasets[21,30] —where models predict materials uploaded year after training—demonstrates that decoupled transfer learning achieves 68% error reduction (RMSE: 0.881 vs. 2.778eV/atom, more than threefold improvement) compared to end-to-end GNNs for extrapolative predictions, while maintaining competitive interpolative performance.

Ablation studies reveal the mechanistic origin of this capability: both components are essential and synergistic—neither pretrained features alone nor simple regressors alone achieve comparable extrapolation. Systematic failure analysis establishes when extrapolation works and when it fails, revealing a fundamental distribution: extrapolation succeeds for continuous extension of training chemical and electric spaces (e.g., more extreme property values within familiar bonding motifs) but encounters difficulties in two scenarios: (i) sparse elemental representation where target elements lack sufficient training examples, and (ii) discontinuous electronic structure transitions involving bonding motifs rarely represented in materials databases. Importantly, the latter reflects the inherent rarity of certain electronic configurations rather than a method limitation—incorporating targeted training examples substantially improves extrapolation in these cases. Successful application to Fermi energy prediction further confirms generalizability beyond formation energy to other density functional theory properties, establishing broad applicability.

These findings transform the paradigm for ML-driven materials discovery. Rather than pursuing ever-larger end-to-end models, the community should prioritize: (1) large-scale pretraining on diverse structures, (2) simple, extrapolation-capable prediction heads, and (3) strategic curation of downstream training data. Critically, this framework is immediately deployable— researchers can use existing pretrained models with standard regression tools, requiring no architectural innovations or extensive computational resources. This provides a practical pathway to accelerate discovery of high-performance materials across energy storage, catalysis, and sustainable technologies.

## Results
### Overview of Approach

Extrapolative property prediction requires two fundamental capabilities: (1) learning generalizable structural representations that transfer across chemical spaces, and (2) extending predictions beyond observed property ranges in training. Conventional end-to-end neural networks fail at extrapolation because joint optimization of feature extraction and property prediction creates coupling between learned representations and target property distributions, constraining model outputs within training ranges.

Our approach addresses this limitation through decoupled transfer learning as presented in Figure 1. We separate representation learning from property prediction by combining pretrained graph neural networks (GNNs) with simple regression models. Specifically, three complementary GNN architectures—CGCNN[12], SchNet[11], and DimeNet++[14] —pretrained on Open Catalyst Project (OC20) dataset provide structural feature vectors. These features are concatenated, normalized, and fed into support vector regression (SVR) or Ridge regression to predict target properties.

This decoupling provides two critical advantages. First, pretrained GNNs encode generalizable structural knowledge—coordination environments, bonding patterns, and geometric motifs learned from millions of diverse structures, enabling meaningful representations even for compositions absent from pretraining data. Second, simple regressors naturally extrapolate through weighted linear combinations of features, extending predictions beyond observed property ranges without the output constraints inherent to deep neural networks. The pretrained representations remain fixed during downstream training, preventing corruption of transferable knowledge while allowing the regression layer to learn property-specific relationships. While we primarily present results using three GNN architectures combined with one SVR, we provide comprehensive ablation studies in the Supplementary Materials examining alternative combinations including single GNN and pairwise GNN combinations (Supplementary Figures 1-3) and Ridge regression variants (Supplementary Figures 4 and 5).

Dataset splitting strategy critically determines whether evaluation reflects actual deployment conditions[20,26]. Thus, rigorous evaluation of extrapolation requires datasets that explicitly separate training and test distributions along meaningful dimensions. Random data splits, while validating interpolative accuracy, fail to assess extrapolation capability because training and test sets share similar compositional, structural, and property distributions. To address this critical methodological gap, we employ two complementary benchmark datasets with deliberate distribution shifts:

**(1) Layered intercalation compounds (LIC) dataset**[28,29]**:** This curated dataset contains 9,024 structures systematically organized into 188 intercalant-erased host structures and 48 intercalant elements. This explicit structure enables controlled evaluation across four distinct scenarios (Figure2), which systematically distinguishes interpolation from three distinct types of extrapolation: (a) random split (interpolation baseline), (b) host-based split (structural extrapolation to novel host frameworks), (c) energy threshold split (property extrapolation to extreme formation energies), and (d) combined split (simultaneous structural and property extrapolation). These scenarios directly reflect realistic materials discovery challenges where predictions must extend beyond training distributions.
This explicit host-intercalant organization provides clear, unambiguous grouping for principled dataset splitting, ensuring rigorous evaluation without data leakage across interpolation and extrapolation scenarios.

**(2) Temporal MP18→MP21 benchmark**[21,30]: This dataset represents real-world temporal extrapolation, where models trained on alloys in the Materials Project 2018 release predict properties of materials added by 2021. This split captures temporal unavailability (future data) and extreme property extrapolation, as MP21 contains substantially more unstable structures than the predominantly stable MP18 training set (MP18: $E_{form}$ = -1.090 to 1.575 eV/atom; MP21: $E_{form}$ = -0.953 to 4.416 eV/atom). Unlike the LIC dataset's controlled structural groupings, this benchmark spans diverse crystal structure types with substantial compositional heterogeneity.

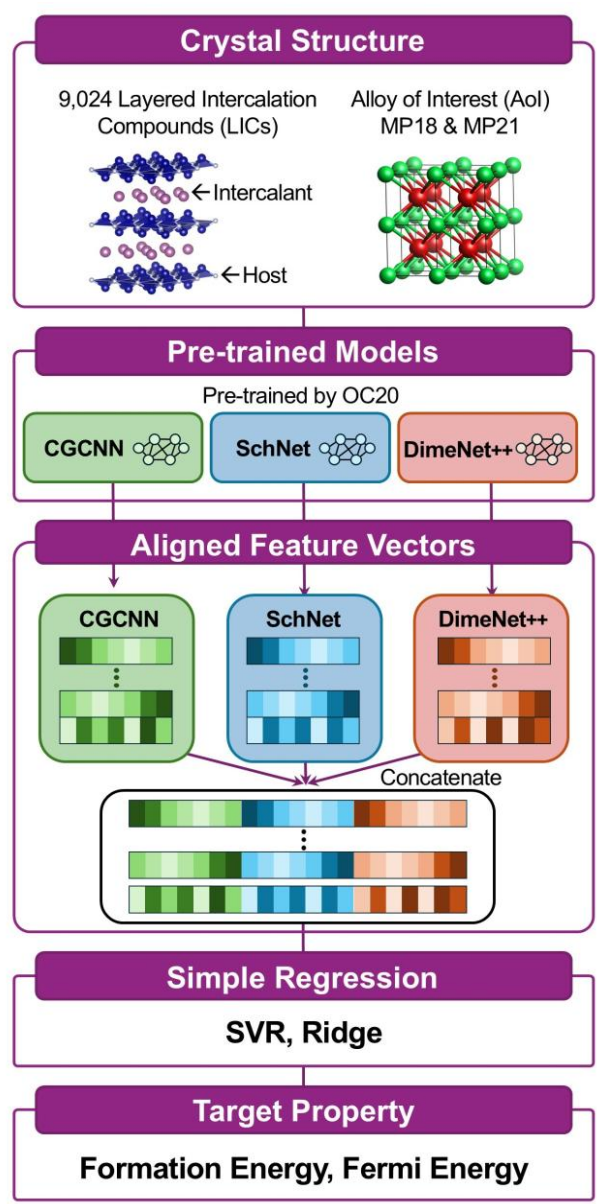

**Figure 1 | Decoupled transfer learning framework combining pretrained GNN feature extractors with simple regression models.** Crystal structures from two benchmark datasets (9,024 layered intercalation compounds and temporal Materials Project alloys, MP18→MP21) are processed through three pretrained GNN architectures (CGCNN, SchNet, and DimeNet++), each capturing complementary structural information. Extracted feature vectors are concatenated and fed into a simple regressor (SVR or Ridge regression) to predict target properties. This decoupled design separates representation learning—handled by frozen pretrained GNNs—from property prediction—handled by simple models with inherent extrapolation capability—enabling genuine extrapolation in both chemical and property space.

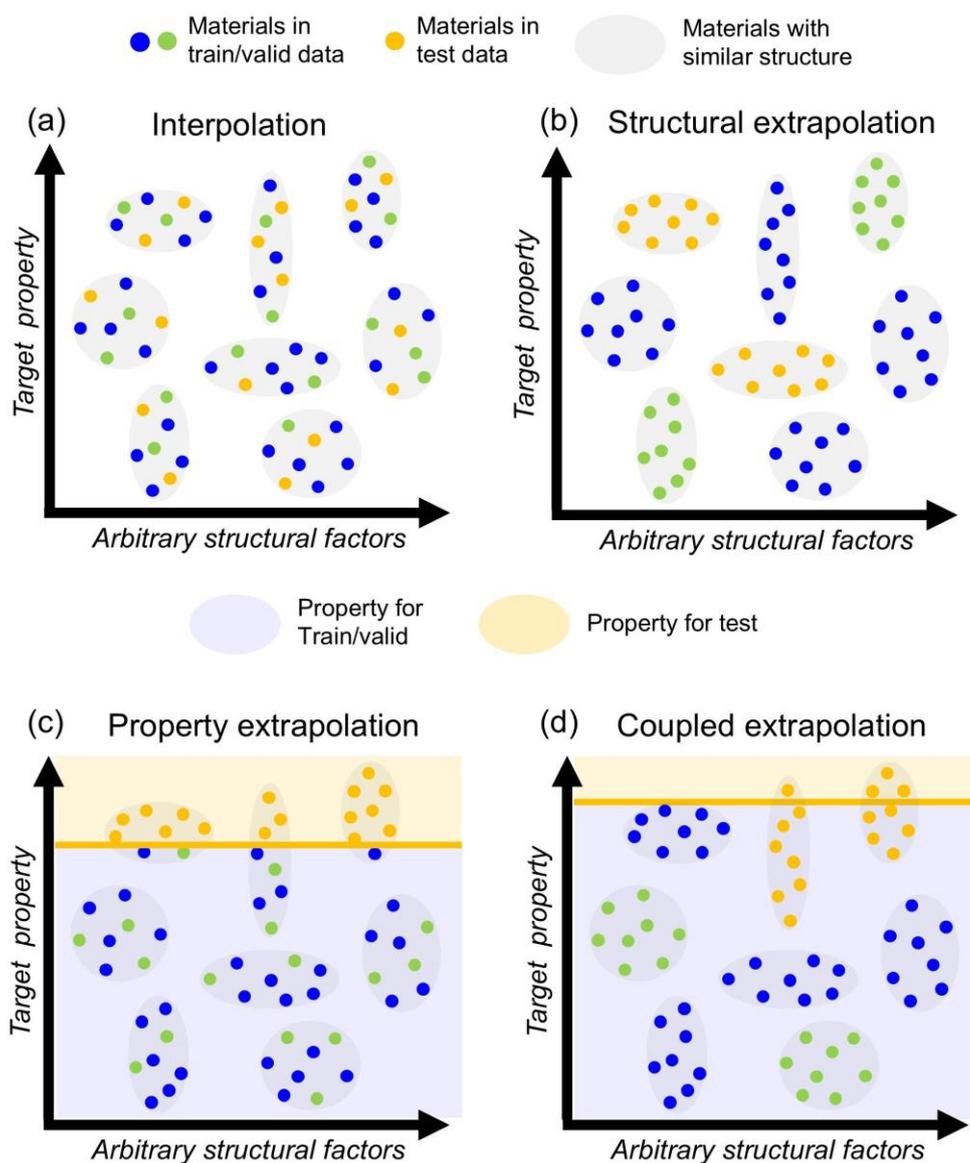

**Figure 2 | Data splitting strategies for systematic evaluation of interpolation and extrapolation performance.** Four scenarios enabled by the explicit host-intercalant structure of the LIC dataset. Blue/green circles represent training/validation materials; orange circles represent test materials. Vertical and horizontal axes represent target property and arbitrary structural factors, respectively. (a) Random split (interpolation baseline): test materials share chemical and property distributions with training data. (b) Host-based split (structural extrapolation): test materials belong to entirely different host structural groups, simulating prediction for novel crystal frameworks. (c) Energy threshold split (property extrapolation): test materials exhibit formation energies outside the training range, simulating the search for compounds with superior or inferior stability. (d) Combined split (coupled extrapolation): test materials differ simultaneously in both structure and property, representing the most challenging and realistic discovery scenario.

We first set aside test sets that represent each intended extrapolation scenario (novel hosts, extreme properties, or temporal data). The remaining data are split into training and validation sets for hyperparameter tuning. This ensures that validation performance, used for model selection, does not leak information about the extrapolation challenge in the test set. Detailed splitting procedures and dataset statistics are provided in Methods.

For comparison, we fine-tuned CGCNN end-to-end on identical data splits to establish baseline performance, enabling direct assessment of how decoupled transfer learning affects extrapolation capability.

**Interpolation Performance establishes baseline capability**

We first verify that our framework doesn't sacrifice accuracy for extrapolation capability—a critical requirement for practical deployment. Under random splitting, where training and test materials share similar compositional and structural distributions, our framework achieves excellent accuracy (Figure 3a). Both out-of-fold validation and test predictions achieve $R^2$ exceeding 0.995, with RMSE of 0.055 eV/atom (test) (Table 1). These results validate the framework's interpolative capability matches end-to-end CGCNN methods (Table 1 and Supplementary Figure 6a).

Remarkably, this high accuracy emerges despite using GNNs pretrained on the Open Catalyst 2020 dataset — small-molecule adsorption on surfaces—which differs substantially from bulk layered intercalation compounds in both chemistry and structural motifs. This cross-domain performance provides strong evidence for the transferability of learned structural representations, supporting our hypothesis that pretrained GNNs encode generalizable geometric and coordination patterns applicable across diverse materials classes.

**Structural Extrapolation: Predicting materials with unseen host structures**

The real test comes with structural extrapolation (Figure 3b). Here, we ask: Can the model predict materials with host structures entirely absent from training? Host-based splitting presents this fundamental challenge: predicting formation energies for materials containing host structures never seen during training (Figure 3b). Test materials may share intercalant elements with training examples but combine them with unseen host frameworks, requiring the model to extrapolate structural knowledge to novel lattice geometries. Predictions generally follow the diagonal trend, indicating successful capture of formation energy trends for unseen hosts, with test RMSE of 0.099 eV/atom (Table 1). Notable outliers including $Y_2O_2$ and graphite compounds, pointed out by circles, represent discontinuous extrapolation cases, analyzed in detail in the failure mode discussion below.

Comparison with end-to-end fine-tuned CGCNN reveal (Table 1 and Supplementary Figure 6(b)) that decoupled transfer learning maintains competitive performance in this structural extrapolation task. While end-to-end CGCNN achieves test RMSE of 0.120 eV/atom, our framework maintains superior performance (0.099 eV/atom), suggesting that pretrained representations transfer more effectively when decoupled from property-specific training. This demonstrates that separating feature learning from property prediction better preserves the structural knowledge needed for extrapolating to novel host frameworks.

**Property Extrapolation: Predicting materials with extreme property values**

Property extrapolation addresses the core challenge of materials discovery: can models predict materials with formation energies far outside the training range? To evaluate this capability, we constructed an energy threshold split where training data are restricted to materials with formation energies between -1.7 and 0.1 eV/atom, while test materials (orange points in Figure 3c) extend beyond these boundaries, representing either highly stable ($E_{form}$ < -1.7 eV/atom) or unstable ($E_{form}$ > 0.1 eV/atom) structures. This scenario directly addresses identifying compounds with unprecedented stability or deliberately targeting metastable phases. The majority of test predictions align with the diagonal, achieving test RMSE of 0.205 eV/atom (Table 1). Critically, our model successfully outputs values beyond training boundaries, with predictions extending to formation energies well outside the -1.7 to 0.1 eV/atom training range.

Comparison with end-to-end CGCNN reveals why conventional approaches fail, RMSE = 0.378 eV/atom

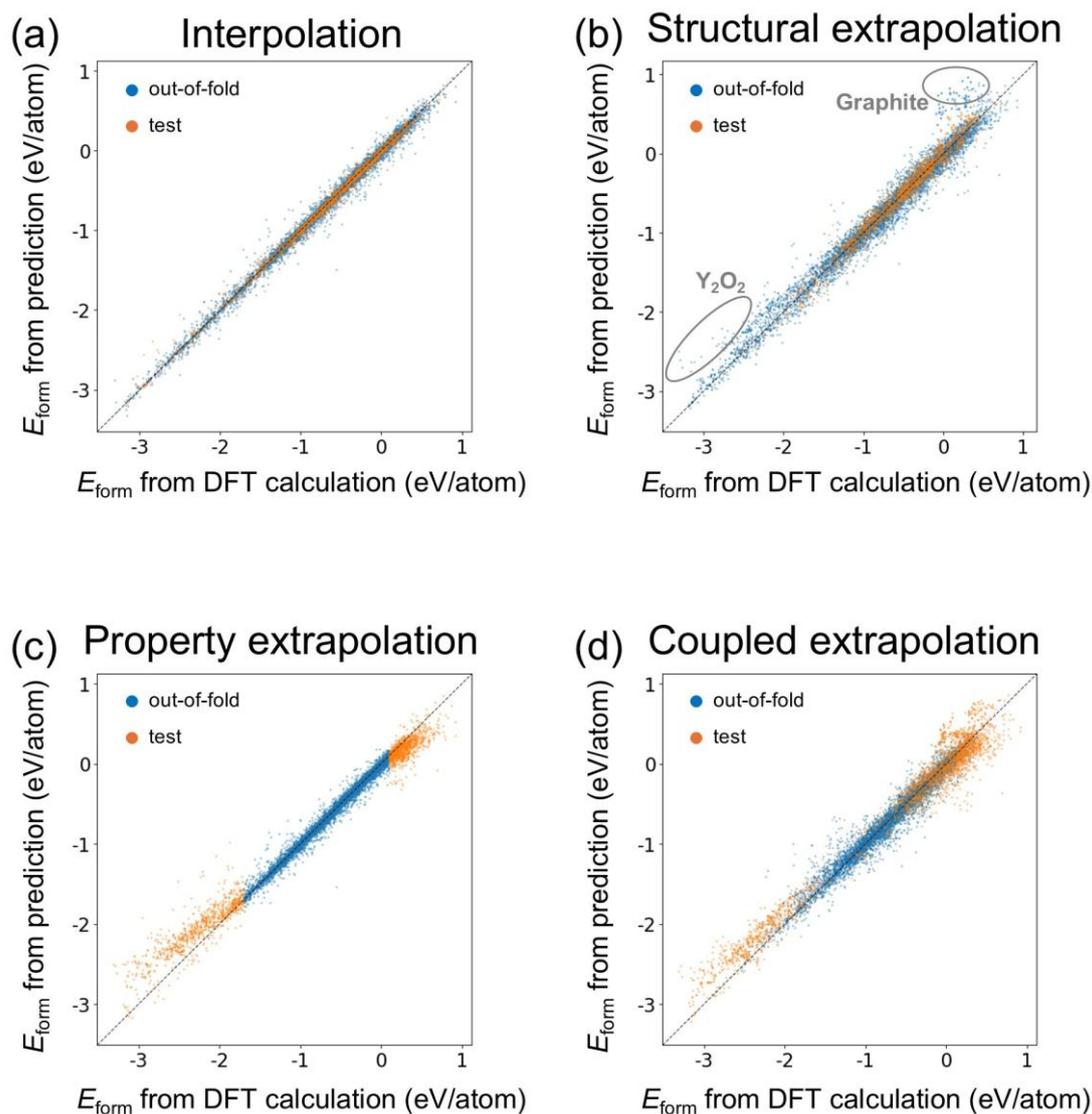

**Figure 3 | Extrapolation performance across multiple scenarios for layered intercalation compounds.** Predicted versus DFT-calculated formation energies (eV/atom) using the decoupled transfer learning approach (3GNNs+SVR) under four splitting strategies. Blue dots: out-of-fold validation; orange dots: test set; gray dashed line: perfect prediction. (a) Interpolation baseline: random split achieves excellent accuracy (test RMSE = 0.055 eV/atom, $R^2$ = 0.995). (b) Structural extrapolation: host-based split demonstrates successful prediction for unseen host structures (test RMSE = 0.099 eV/atom, $R^2$ = 0.969); circled regions highlight systematic outliers from $Y_2O_2$-based compounds (sparse elemental representation) and graphite-based compounds (discontinuous electronic structure). (c) Property extrapolation: energy threshold split shows robust extrapolation beyond training boundaries (test RMSE = 0.205 eV/atom, $R^2$ = 0.975), with test predictions successfully extending outside the training range. (d) Coupled extrapolation: combined split maintains reasonable accuracy (test RMSE = 0.199 eV/atom, $R^2$ = 0.966) under simultaneous structural and property extrapolation. Quantitative comparison with end-to-end CGCNN is provided in Table 1 and Supplementary Fig. 6.

(Table 1 and Supplementary figure 6(c)). CGCNN predictions cluster near the training range, failing to extrapolate beyond observed values—a characteristic limitation of deep neural networks trained end-to-end. In contrast, our approach produces outputs spanning the full test range, demonstrating the inherent extrapolative capability of simple regressors when provided with meaningful structural features. This performance gap directly validates our core hypothesis: decoupling representation learning from property prediction enables genuine extrapolation.

**Coupled Structural and Property Extrapolation**
Combined splitting represents the most challenging materials discovery scenario: simultaneous structural and property extrapolation (Figure 3d). In this case, when a group includes at least one structure with a formation energy above the threshold, the entire group is designated as the test set., reflecting the challenge of finding unprecedented structures with superior properties. Despite this difficulty, predictions maintain reasonable agreement with the diagonal (test RMSE = 0.199 eV/atom). The majority of test points are correctly positioned at the orthogonal line, indicating that the model captures formation energy trends even when both structural and property extrapolation are required simultaneously.

As in property extrapolation, our method substantially outperforms end-to-end CGCNN, which fails to produce outputs beyond its training distribution (Table 1 and Supplementary Figure 6(d)), RMSE = 0.308 eV/atom. This performance demonstrates that decoupled transfer learning provides robust extrapolation even under the most demanding conditions where multiple distribution shifts occur simultaneously.

**Table 1 | Quantitative comparison of prediction performance across different scenarios on the layered intercalation compound (LIC) dataset.**
The decoupled transfer learning approach (3GNNs+SVR) outperforms end-to-end fine-tuned CGCNN across all scenarios: interpolation (RMSE < 0.07 eV/atom for both, confirming competitive baseline performance), structural extrapolation (18% RMSE reduction: 0.099 vs. 0.120 eV/atom), property extrapolation (46% RMSE reduction: 0.205 vs. 0.378 eV/atom), and coupled extrapolation (35% RMSE reduction: 0.199 vs. 0.308 eV/atom). Underlined values indicate superior performance. MAE: mean absolute error; RMSE: root mean square error; $R^2$: coefficient of determination.

| | Interpolation | | | Structural extrapolation | | | Property extrapolation | | | Coupled extrapolation | | |
|---|---|---|---|---|---|---|---|---|---|---|---|---|
| | MAE (eV/atom) | RMSE (eV/atom) | $R^2$ | MAE (eV/atom) | RMSE (eV/atom) | $R^2$ | MAE (eV/atom) | RMSE (eV/atom) | $R^2$ | MAE (eV/atom) | RMSE (eV/atom) | $R^2$ |
| **Proposed method (3GNNs+SVR)** | 0.035 | 0.055 | 0.995 | 0.076 | 0.099 | 0.969 | 0.150 | 0.205 | 0.975 | 0.155 | 0.199 | 0.966 |
| **End-to-end CGCNN** | 0.045 | 0.063 | 0.993 | 0.093 | 0.120 | 0.955 | 0.281 | 0.378 | 0.914 | 0.237 | 0.308 | 0.918 |

Having systematically evaluated extrapolation across controlled scenarios, we now benchmark our framework against a real-world temporal extrapolation challenge where principled data splitting is infeasible— the Materials Project alloy dataset (MP18→MP21). This temporal benchmark represents the ultimate test of extrapolative capability under actual materials discovery conditions.

**Temporal Extrapolation Benchmark**
The MP18→MP21 temporal benchmark provides real-world validation of extrapolation capability (Figure 4 and Table 2). Models trained on alloy structures available in the Materials Project 2018 release predict properties of materials added by 2021 —representing genuine temporal extrapolation to unavailable future data[21]. This split simultaneously presents extreme property extrapolation: training and validation data span -1.090 to 1.575 eV/atom, while test materials extend to 4.416 eV/atom, encompassing highly unstable structures rarely observed in MP18.

Our method achieves remarkable performance (Figure 4): test RMSE = 0.296 eV/atom, $R^2$ = 0.873. While scatter increases for highly unstable structures ($E_{form}$ > 1.58 eV/atom), predictions maintain qualitatively correct trends even up to 4.416 eV/atom. The model still captures relative stability rankings, enabling meaningful screening of unstable phases.

Table 2 compares our results with prior methods evaluated on identical data. The best previous approach (XGBoost with Matminer features) achieved RMSE = 0.537 eV/atom for the all data including interpolative and extrapolative data, representing 45% RMSE improvement by using the present decoupled transfer learning method. However, this overall metric —which mixes both interpolative and extrapolative data— obscures critical differences in extrapolation capability. End-to-end fine-tuned CGCNN demonstrates the fundamental limitation of coupled training. In the interpolation region (between -1.090 to 1.575eV/atom), CGCNN achieves RMSE = 0.332 eV/atom — reasonable performance when predicting within the training distribution (Table 2). However, performance catastrophically degrades in the property extrapolation region (above 1.575eV/atom): RMSE = 2.778 eV/atom, more than threefold worse. This collapse reflects the inherent constraint of end-to-end neural networks, which struggle to produce outputs beyond observed training ranges. In contrast, the present decoupled transfer learning maintains robust performance across both regions. Our approach, 3GNNs-SVR, achieves RMSE = 0.151 eV/atom (interpolation) and RMSE = 0.881 eV/atom (extrapolation), representing 55% improvement over end-to-end CGCNN in the interpolation region and critically, 68% improvement in the extrapolation region. This dramatic performance gap in extrapolation validates our core design principle: separating representation learning from property prediction enables genuine extension beyond training distributions while preserving interpolative accuracy.

Moreover, while decision-tree-based method, such as XGBoost with Matmainer features, fundamentally failed to extrapolate beyond moderate formation energies[21], our approach maintains meaningful predictions across the full spectrum to 4.416 eV/atom, including highly unstable materials. This demonstrates that decoupled transfer learning overcomes the extrapolation barrier that limits both end-to-end neural networks and traditional machine learning approaches.

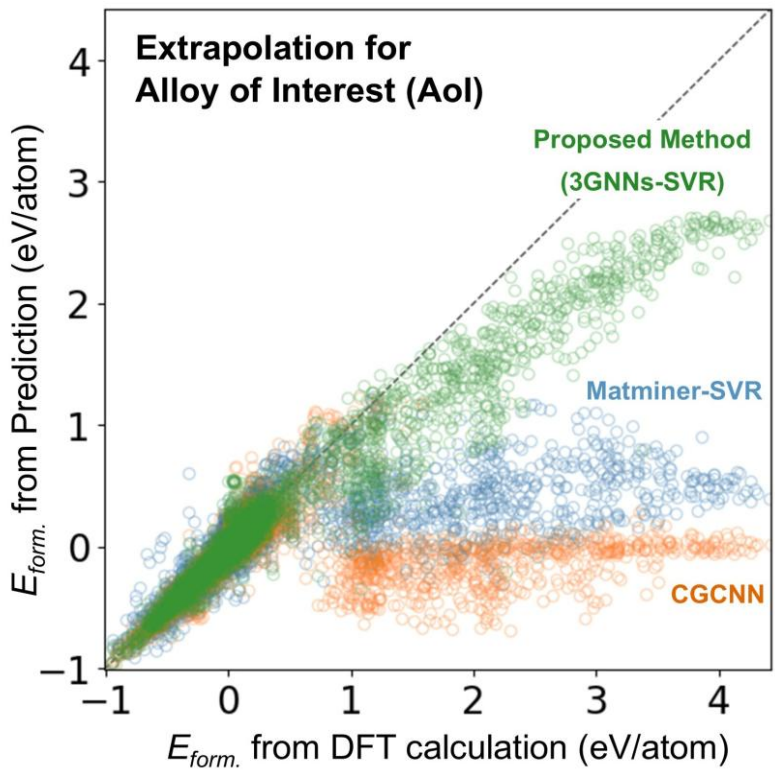

**Figure 4 | Temporal extrapolation benchmark on the Materials Project alloy dataset (MP18→MP21).** Predicted versus DFT-calculated formation energies (eV/atom) for three approaches: proposed decoupled transfer learning (3GNNs+SVR, green), Matminer-SVR (blue), and end-to-end fine-tuned CGCNN (orange). Gray diagonal line indicates perfect prediction. The vertical dashed line at $E_{form}$= 1.575 eV/atom separates the interpolation region (within MP18 training range) from the extrapolation region (extending to 4.416 eV/atom in MP21). Our method achieves RMSE = 0.151 eV/atom (interpolation) and 0.881 eV/atom (extrapolation), representing 68% error reduction over end-to-end CGCNN (RMSE = 2.778 eV/atom) in the extrapolation region. Matminer-SVR achieves extrapolation RMSE = 2.262 eV/atom, confirming that simple regressors alone are insufficient without rich pretrained GNN features. Complete visualization including out-of-fold predictions is provided in Supplementary Fig. 7.

**Table 2 | Performance comparison on the temporal Materials Project benchmark (MP18→MP21) demonstrating real-world extrapolation capability.** Interpolation and extrapolation regions are separated at $E_{form}$ = 1.575 eV/atom, corresponding to the MP18 training range (−1.090 to 1.575 eV/atom) and the MP21 extrapolation range (1.575 to 4.416 eV/atom), respectively. The decoupled approach (3GNNs+SVR) achieves 68% RMSE reduction in the extrapolation region (0.881 vs. 2.778 eV/atom) compared to end-to-end CGCNN, and outperforms all prior methods including ALIGNN, XGBoost, Random Forest, and Linear Forest. Matminer-SVR (extrapolation RMSE = 2.262 eV/atom) confirms that rich pretrained GNN representations are essential even when a simple regressor is used. Underlined values indicate superior performance.

|  | Models | MAE (eV/atom) | RMSE (eV/atom) | $R^2$ |
|---|---|---|---|---|
| **This study** | Proposed method (3GNNs-SVR) | 0.143 | Overall: 0.296<br>Interpolation: 0.151<br>Extrapolation: 0.881 | 0.873 |
|  | End-to-end CGCNN | 0.356 | Overall: 0.876<br>Interpolation: 0.332<br>Extrapolation: 2.778 | -0.112 |
|  | Matminer-SVR | 0.297 | Overall: 0.702<br>Interpolation: 0.240<br>Extrapolation: 2.262 | 0.284 |
| **Previous study[21]** | ALIGNN | 0.297 | 0.747 | 0.194 |
|  | XGBoost | 0.239 | 0.537 | 0.582 |
|  | Random Forest | 0.382 | 0.879 | -0.119 |
|  | Linear Forest | 0.327 | 0.606 | 0.469 |

## Discussion
### Ablation Studies and Mechanistic Understanding
To elucidate the source of extrapolative capability, we conducted systematic ablation studies decomposing our framework into its constituent components (Table 2 and Figure 4). These experiments reveal that neither pretrained features nor simple regressors alone achieve comparable performance—both components are essential and synergistic. End-to-end fine-tuned CGCNN performed poorly (RMSE = 0.876 eV/atom, $R^2$ = -0.112 for overall region), with predictions collapsing toward the training distribution. This poor performance, despite CGCNN's architectural sophistication, confirms that end-to-end optimization fundamentally limits extrapolation capability by coupling learned representations to target property distributions. Simple regression with Matminer descriptors (Matminer-SVR) achieves RMSE = 0.702 eV/atom, $R^2$=0.284 for overall region, substantially better than end-to-end CGCNN but still inferior to our full framework (RMSE=0.296eV/atom, $R^2$=0.873, Figure 4, Table 2, and Supplementary Figure 7).

This result demonstrates two critical insights: First, simple regressors provide inherent extrapolation capability compared to deep neural networks, validating the second component of our design. Second, Matminer descriptors, despite incorporating domain knowledge, cannot match the richness of representations learned by GNNs from millions of diverse structures. Features matter—even with an extrapolation-capable regressor, inadequate structural representations limit performance.

Using pretrained GNN features with SVR (our full approach) achieves RMSE = 0.296 eV/atom for the alloy dataset, 68% improvement over end-to-end CGCNN. This dramatic performance gain emerges only when both components combine: transferable structural representations from pretrained GNNs AND extrapolation-capable simple regression.

Further ablation examining the number of pretrained GNNs (n = 1, 2, 3) reveals that ensemble diversity enhances performance (Supplementary Figures 1 and 2 for LIC and Supplementary figure 3 for alloy). Individual GNN architectures (CGCNN, SchNet, or DimeNet++) provide complementary structural perspectives—CGCNN emphasizes local coordination, SchNet captures radial distributions, and DimeNet++ encodes angular information. Combining all three yields the most robust and generalizable feature representation.

Mechanistic understanding of this success reflects complementary strengths. Pretrained GNNs learn generalizable structural representations from diverse materials databases, encoding geometric patterns, coordination environments, and electronic structure signatures that transfer across chemical spaces. Critically, pretraining on millions of structures spanning diverse compositions ensures that the model encounters varied bonding motifs, coordination numbers, and lattice geometries. Although OC20 focuses on surface adsorption while our targets involve bulk formation energies, the learned features encode transferable structural knowledge applicable to different chemical contexts. This broad foundation enables meaningful representations even for compositions absent from pretraining data, provided structural motifs remain recognizable.

The role of the simple regressor is also important. Results using another simple regressor, Ridge, for the LIC dataset are shown in Supplementary Figures 4 and 5 for LIC and alloy, respectively. The performance is slightly inferior to the model using SVR, particular for coupled extrapolation, but extrapolation performance remains much better than the end-to-end CGCNN method. This confirms that simple regressors provide the complementary capability that deep neural networks lack: natural extrapolation through weighted linear combinations of learned features. Deep neural networks, when trained end-to-end, produce outputs constrained by training distributions because the property prediction layers learn to map features to observed value ranges. In contrast, SVR and Ridge extrapolate inherently—given features encoding structural variations, linear combinations naturally extend beyond observed property ranges.

This synergy—pretrained GNNs encode "what makes structures different" (geometric and chemical variations), while simple regressors learn "how these differences affect properties" through transferable linear relationships—is critical. Decoupling these functions prevents the representation corruption that occurs in end-to-end training, where optimization pressure from limited target property ranges distorts learned structural features.

**Failure mode analysis reveals extrapolation boundaries**

Understanding when and why extrapolation fails is as important as demonstrating success. Analysis of outliers in structural extrapolation (Figure 3b) identifies compounds with $Y_2O_2$ and graphite hosts exhibiting systematically larger prediction errors, as pointed out by circles. These failures reveal a fundamental distinction between continuous extrapolation and discontinuous extrapolation, with each failure mode stemming from different causes of discontinuity.

**Continuous extrapolation** extends predictions along established trends within known chemical regimes. Test materials differ quantitatively from training data—more extreme property values, novel element combinations within similar chemical families or new structural arrangements of familiar motifs—but the underlying physics remains consistent. Our framework excels at continuous extrapolation: predicting extrapolative formation energies of LIC (below -1.7eV/atom and over 0.1eV/atom) from training data spanning -1.7 to 0.1 eV/atom (Figure 3(c)).

**Discontinuous extrapolation** involves qualitative jumps to distinct chemical regimes with fundamentally different bonding, electronic structures, or coordination environments poorly represented in training data. These present inherent challenges because the model lacks sufficient examples of the structural patterns governing properties in these regimes.

Our failure mode analysis identifies two instructive examples of discontinuous extrapolation.

**Sparse task-relevant elemental representation: The $Y_2O_2$ case**
$Y_2O_2$-based compounds reveal a critical requirement for extrapolation: adequate task-relevant training data. Yttrium appears in the OC20 pretraining dataset (among 55 substrate elements), yet $Y_2O_2$ intercalation compounds exhibit elevated prediction errors. The issue is not elemental absence from pretraining but scarcity in the downstream task: our layered intercalation dataset contains very few Y-containing structures. During structural extrapolation evaluation where $Y_2O_2$ hosts are excluded from training, the model lacks sufficient examples to learn how yttrium-containing structures relate to formation energies, creating a discontinuous jump to an underrepresented element.

This contrasts with lithium, abundant in our intercalation dataset but absent from OC20 pretraining. Despite lacking pretraining exposure, the model successfully predicts Li-containing compounds because sufficient downstream training examples enable learning Li's formation energy behavior. Similarly, elements like Be, Mg, Ba, Fr, Ra, F, Br, and I—present in the intercalation dataset but absent from OC20—are accurately predicted when adequate training examples exist.

This demonstrates a key principle: downstream task learning compensates for pretraining absence, but the reverse fails. Pretraining on different contexts (adsorption energies) provides limited transferability to qualitatively different properties (bulk formation energies) when downstream examples are sparse. Successful extrapolation requires either (1) sufficient target element examples in downstream training data, enabling task-specific learning, or (2) high chemical similarity to well-represented elements, enabling transfer across similar coordination environments. $Y_2O_2$ satisfies neither: sparse downstream examples and limited geometric similarity to abundant elements in the intercalation dataset.

**Electronic structure discontinuity: The graphite case**
Graphite-based compounds present a complementary failure mode, as circled in Figure 3b: discontinuity from rare electronic configurations rather than elemental scarcity. Carbon is well-represented in both OC20 and our intercalation training set, yet graphite intercalation compounds exhibit elevated prediction errors.

The issue is electronic structure distinctiveness. Graphite is the only neutral host in our dataset, featuring $sp^2$ hybridization with extended π-electron delocalization. This contrasts sharply with the ionic hosts dominating the training set of layered intercalation compounds, which exhibit localized bonding and charge transfer. Critically, neutral $sp^2$ π-systems constitute an extremely small fraction of stable bulk materials in chemical space—most crystalline compounds are ionic or metallic. Although both OC20 (C-based substrates and molecules) and our training set (carbides) provide carbon-related information, neither adequately represents this rare electronic regime.

This contrasts with the $Y_2O_2$ case, where failure stemmed from insufficient downstream examples. Here, carbon is abundant but graphite-like electronic character is rare. Linear combinations of learned features struggle to capture qualitative electronic transitions to states existing in only a tiny fraction of stable compounds. This represents discontinuous extrapolation: a discrete jump to an underrepresented electronic structure regime rather than a continuous extension within familiar bonding motifs.

This observation reframes certain extrapolation challenges as reflecting natural scarcity in chemical space rather than algorithmic limitations. For graphite, the issue is not inadequate elemental coverage but the statistical rarity of extended π-systems among bulk intercalation hosts. Addressing such cases requires either (1) explicit inclusion of rare electronic motifs in training data when targeting these regimes, or (2) physics-informed constraints encoding relevant electronic structure differences.

**Design principles for continuous extrapolation.**
These contrasting failure modes establish refined design principles:
**(1) Prioritize task-relevant elemental coverage.** Pretraining provides transferable geometric representations, but element-specific property relationships require sufficient downstream training examples. Lithium's success—absent from pretraining but abundant in training—demonstrates that downstream learning compensates for pretraining gaps.
**(2) Recognize task similarity limits.** Pretraining on different property types (adsorption versus bulk formation energies) provides primarily structural rather than energetic transferability. When pretraining and target tasks involve qualitatively different properties, downstream examples become critical for element-specific learning.
**(3) Assess electronic structure continuity.** Extrapolation succeeds for continuous extensions of training electronic configurations but struggles with discrete transitions—ionic to neutral, localized to delocalized bonding—involving configurations statistically rare in the target materials class. The graphite case (abundant element, rare electronic state) complements the $Y_2O_2$ case (sparse element, typical electronic state) in defining extrapolation boundaries.
**(4) Curate training data strategically.** For underrepresented elements, include diverse downstream training examples rather than relying solely on pretraining. For well-represented elements with rare electronic structures (novel bonding motifs), incorporate representative examples to establish these regimes within the model's learned chemical space, converting discontinuous extrapolation into continuous extension.

**Generalization Beyond Formation Energy**

To assess whether decoupled transfer learning applies broadly or remains specific to formation energy, we evaluated the framework on Fermi energy prediction—an electronic property with fundamentally different physical character. Formation energy reflects thermodynamic stability through total energy differences, while Fermi energy characterizes electronic structure through occupied state distributions. Applying identical methodology to DFT-calculated Fermi energies demonstrates comparable extrapolation capability across all four splitting scenarios (Supplementary Figures 8-10, Supplementary Table 1). Interpolation achieves test RMSE = 0.352 eV, structural extrapolation yields RMSE = 0.502 eV, property extrapolation maintains RMSE = 1.269 eV despite targeting values outside training ranges, and coupled extrapolation achieves RMSE = 1.200 eV. The relative performance pattern mirrors formation energy results: successful extrapolation in both structural and property space, substantially outperforming end-to-end GNN baselines (Supplementary Figure 10 and Supplementary Table 1).

These results establish that decoupled transfer learning extends beyond formation energy to other single-structure DFT-computable properties. The consistent extrapolation patterns across both thermodynamic (formation energy) and electronic (Fermi energy) properties suggest broad applicability within this class of materials descriptors. However, properties requiring fundamentally different calculations—ensemble averaging (polycrystalline properties), dynamic processes (ionic conductivity, diffusion), or explicit defects—may present additional challenges requiring methodological extensions. The demonstrated success for two physically distinct single-structure properties provides strong evidence for generalizability while establishing clear scope boundaries.

**Impact on materials discovery paradigm**

Our findings fundamentally reshape the trajectory of ML-driven materials discovery. The persistent failure of sophisticated end-to-end models at extrapolation—despite years of architectural innovations—reveals that complexity is not the solution. Instead, this work establishes a counterintuitive principle: simplicity enables extrapolation. Decoupled transfer learning achieves 68% error reduction (RMSE: 0.881 vs. 2.778 eV/atom) in extrapolative predictions compared to the end-to-end methods, while matching their interpolative accuracy. This performance breakthrough emerges not from novel architectures but from strategic separation of representation learning and property prediction.

The practical implications are immediate and transformative. Researchers can deploy this framework today using existing pretrained GNN models (such as CGCNN, SchNet, and DimeNet++) combined with standard regression tools—no architectural innovations, no extensive computational resources, no specialized expertise required. This democratizes access to robust extrapolative prediction, previously unachievable even with cutting-edge methods. For battery materials, catalysts, and energy storage systems, this enables confident prediction of compounds with unprecedented stability or performance, accelerating the discovery cycle from computational screening to experimental validation.

Looking forward, our failure mode analysis provides actionable design principles. Continuous extrapolation succeeds when targeting more extreme property values within familiar bonding motifs or novel element combinations within similar chemical families. Strategic data curation—prioritizing task-relevant elemental coverage and incorporating representative examples of rare electronic configurations—converts challenging discontinuous extrapolation into tractable continuous extension. Combined with the established four-scenario evaluation framework (interpolation, structural extrapolation, property extrapolation, coupled extrapolation), these principles enable systematic assessment of extrapolation reliability before experimental investment.

This work resolves the long-standing trade-off between interpolative accuracy and extrapolative capability, providing a robust framework and clear methodology for reliable ML-driven materials discovery. By demonstrating generalizability across formation energy and Fermi energy prediction, we establish that decoupled transfer learning extends broadly to single-structure DFT properties. These findings provide practical guidelines for applying machine learning to identify genuinely novel, high-performance materials—advancing the practical realization of accelerated materials discovery for global challenges in energy, catalysis, and sustainable technologies.

**Practical Implications and Future Directions**

Our framework provides immediate advantages for materials discovery. The demonstrated 68% extrapolation improvement directly translates to reduced false negative rates in computational screening, potentially decreasing experimental validation efforts and accelerating discovery timelines.

Several directions could further enhance extrapolation capability. Physics-informed feature augmentation—combining GNN representations with domain knowledge (electronegativity, ionic radii, oxidation states) —may improve predictions for elementally sparse or electronically distinct materials while preserving learned structural knowledge. Active learning strategies that identify and selectively augment training data with underrepresented elements or rare electronic configurations could systematically address discontinuous extrapolation challenges. Uncertainty quantification through ensemble methods or Bayesian regression would enable principled identification of extrapolation boundaries, flagging predictions requiring experimental validation. Validation across additional properties (band gaps, mechanical moduli, catalytic descriptors) and material classes would demonstrate broader applicability throughout the materials discovery pipeline.

In summary, decoupled transfer learning—combining pretrained GNN features with simple regression—enables robust extrapolation in materials property prediction, achieving 68% improvement over end-to-end approaches in extrapolation regions while maintaining competitive interpolative accuracy. Our framework succeeds for continuous extensions of training chemical spaces but encounters fundamental barriers with discontinuous jumps involving sparse task-relevant representation or rare electronic configurations. These findings establish both an immediately deployable tool for computational materials discovery and design principles clarifying when extrapolation is reliable, when targeted data curation is necessary, and when complementary validation is warranted—transforming extrapolation from an unpredictable capability into a systematically manageable aspect of materials informatics.

## Methods
### Proposed Framework and Model Architecture

Our approach combines pretrained GNN feature extractors with simple regression models to achieve robust extrapolation. This design is motivated by two complementary observations. First, GNNs pretrained on large and diverse datasets can extract informative structural representations even for materials outside their training distribution, leveraging broad exposure to varied chemical environments through transfer learning. Second, simple regression models such as linear models and kernel methods like SVR can naturally extrapolate beyond the training property range through linear combinations of features, unlike deep neural networks whose outputs tend to cluster near the training data. By decoupling representation learning—performed by pretrained GNNs rich in structural knowledge—from property prediction—handled by simple models with inherent extrapolative capability—this decoupled transfer learning framework addresses extrapolation in both chemical space (novel materials) and property space (extreme property values), effectively overcoming the accuracy-extrapolation trade-off that limits existing approaches.

Figure 1 illustrates the overall architecture of the proposed decoupled transfer learning framework. In the first stage, input crystal structures are processed through multiple pretrained GNN architectures to extract complementary structural features. These feature vectors are then concatenated and fed into a simple regression model to predict target properties. This decoupled design contrasts with conventional end-to-end fine-tuning of GNNs, which we use as a baseline for comparison.

For feature extraction, we employed three pretrained GNN models validated as effective feature extractors in previous studies[17]: CGCNN[12], SchNet[11], and DimeNet++[14]. All models are pretrained on the OC20 dataset[16], which focuses on small-molecule adsorption on surfaces and contains diverse atomic environments. Importantly, it does not include the bulk crystalline materials used in our downstream tasks, ensuring genuine evaluation of transfer learning capability.

The three architectures capture progressively richer structural information. CGCNN captures bond distances through discrete binning, SchNet processes continuous distance information, and DimeNet++ additionally incorporates bond angle information. All three models are rotationally invariant. We selected this ensemble based on the hypothesis that combining their complementary representations would provide a richer and more general feature space than any single model. Following prior work[17], we extracted atomic-level vector representations from input structures and averaged them to obtain structure-level features. Feature vectors from the three models were concatenated, normalized, and fed into the regression model described below.

For the second-stage regressor, we evaluated Support Vector Regression (SVR) with RBF kernel and Ridge Regression as simple models. Although SVR with RBF kernel is nonlinear in the feature space, it performs linear prediction after kernel transformation, thereby maintaining a certain degree of extrapolative capability in the output space, provided that the distance between the prediction target and training data is not too large. This contrasts with deep neural networks, which tend to produce outputs within their training range. While both methods demonstrated output extrapolation, SVR with RBF kernel was more robust to outliers than Ridge (Ridge results are provided in Supplementary Figures 4 and 5).

For comparison with our decoupled transfer learning approach, we fine-tuned CGCNN, one of the pretrained models, directly on the target datasets in an end-to-end manner. This baseline represents the conventional approach of adapting foundation models to downstream tasks and allows us to assess whether the decoupled transfer learning framework offers advantages for extrapolation.

**Datasets and Splitting Strategies**
To rigorously evaluate extrapolation performance across diverse scenarios, we employed two complementary datasets: (1) a layered intercalation compounds dataset[28,29] enabling systematic evaluation of well-defined extrapolation scenarios through explicit structural grouping, and (2) a Materials Project-derived alloy dataset[21,30] for benchmarking against established temporal extrapolation studies.

Dataset splitting is critically important for evaluating model performance in realistic materials discovery scenarios[20,26]. Without data splits that reflect actual deployment conditions, assessing whether models will function in practice becomes impossible. Accordingly, it is necessary to first set aside a test set that faithfully reflects the intended deployment distribution. The remaining data are then split into training and validation sets to tune hyperparameters. Figure 2 summarizes our dataset splitting strategies.

Figure 2a shows conventional random splitting, which assumes interpolation. Blue and green dots represent materials in training and validation sets, respectively, while orange dots indicate test materials. The vertical axis represents the target property (e.g., formation energy), and the horizontal axis represents an arbitrary dimension capturing structural and compositional diversity. In Figure 2a, blue/green dots surround orange dots—predicting orange from blue/green constitutes interpolative prediction.

In contrast, evaluating material extrapolation (Figure 2b-d) is considerably more challenging. Feature-based clustering is commonly employed[20,26,33], but structural similarity and feature similarity are not equivalent in materials[33]. Moreover, whether data are "known" bears no relation to feature similarity in actual development scenarios. The required extrapolation capability is predicting "data unlikely to be available" from "data at hand." Achieving data splits matching this scenario requires complete understanding of the dataset's characteristics. However, general-purpose datasets like Materials Project contain highly diverse materials, making principled structural grouping impossible. This often leads to data leakage and performance overestimation[20].

Given this background, we primarily employed a layered intercalation compounds dataset [28,29] with well-defined material characteristics amenable to clear splitting. This dataset comprises 9,024 structures derived from 188 host structures and 48 intercalants, along with their first-principles calculation results. Because structures are explicitly constructed from host-intercalant combinations with clear flags, we can perform rigorous grouping with careful attention to data leakage. The target property is formation energy per atom.

To assess both interpolation and various extrapolation scenarios, we implemented four splitting strategies (Figure 2):

(a) **Random split (interpolation, Figure 2a)**: As described above, this assumes predicting properties of materials similar to known ones.

(b) **Host-based split (structural extrapolation, Figure 2b)**: Test materials belong to different structural groups than training/validation materials, while property values for test materials fall within the training range. This corresponds to structural extrapolation—predicting properties of previously unexplored material classes. For layered compounds, this simulates predicting properties for new host structures.

(c) **Energy threshold split (property extrapolation, Figure 2c)**: Test materials belong to the same structural groups as training materials, but their property values lie outside the training range. This represents property extrapolation—predicting performance exceeding existing data for materials near known compositions, simulating the search for superior performance beyond current datasets.

(d) **Combined split (coupled extrapolation, Figure 2d)**: In this case, when a group includes at least one structure with a formation energy above the threshold, the entire group is designated as the test set. This simulates searching for unknown materials exceeding existing material performance—the most challenging and realistic scenario.

For (a), we performed 4:1 training-validation:test random splitting with random validation splits. For (b), we grouped materials by their host composition, ensuring that no host compositions appearing in the test set were present in the training or validation sets (4:1 ratio). The layered intercalation dataset comprises 188 host-intercalant configurations; however, some pairs such as $C_8X_{abc}$ and $C_8X_{abcd}$ share the same composition (graphite) but differ in the stacking sequences, abc and abcd, respectively. To prevent potential data leakage from such structurally related compounds, we grouped data by host composition rather than individual structural configurations, yielding 139 distinct groups for grouped cross-validation. Furthermore, to prevent overfitting through data leakage during validation, we also ensured that the validation and training sets contained no overlapping host compositions. For (c), structures with $E_{form} \geq 0.1$ eV/atom or $\leq -1.7$ eV/atom comprised the test set, with random validation splits. For (d), after host-based grouping, groups containing structures with $E_{form} \geq 0.4$ eV/atom or $\leq -2.6$ eV/atom formed the test set, with validation ensuring no host overlap with training.

For comparison with prior extrapolation studies, we employed the Alloys of Interest (AoI) dataset[21,30], comprising alloy data constructed from the first 34 metallic elements in the Materials Project, using the same data and train/test splits as in previous work. The training/validation set consists of the AoI in MP18 (Materials Project 2018.06.01 version), whereas the test set comprises AoI that appear only in MP21 (Materials Project 2021.11.10 version), thereby evaluating temporal extrapolation by predicting future data. This split simultaneously represents temporal extrapolation (unavailable future data) and significant property extrapolation: MP18 formation energies span -1.090 to 1.575 eV/atom, while MP21 extends to -0.953 to 4.416 eV/atom, including highly unstable structures. Note that principled structural grouping is infeasible for this dataset, preventing clear interpolation vs. structural extrapolation distinction—however, comparison with prior results demonstrates our method's advantages. For consistency with prior work, the target property is formation energy per atom. Random splits were performed for validation. Critically, materials in the AoI dataset do not exist in the OC20 pretraining dataset, which focuses on small-molecule adsorption. Thus, pretrained models have never encountered these specific materials during training.

**Model Training and Hyperparameters**

To ensure reproducibility, all random seeds were fixed, and identical data splits were used in each validation experiment. For the SVR model using the RBF kernel, we employed the cuml.svm.SVR implementation[34], which supports GPU acceleration. We performed 4-fold cross-validation, tuning C and ε via grid search to minimize root mean square error (RMSE) on the out-of-fold (OOF) predictions (C: 10-1000 in 5 steps; ε: 0-0.06 in 0.02 increments). Test predictions used the average of four cross-validated models.

For the comparison GNN baseline, we fine-tuned OC20-pretrained CGCNN[12] on the target datasets using 3:1 holdout validation. Training was conducted with a mini-batch size of 32 using the AdamW optimizer[35]. The initial learning rate was set to $1 \times 10^{-4}$. The learning rate decayed by a factor of 0.5 when the RMSE of the validation dataset did not improve for 2 consecutive epochs. Learning was stopped when the validation RMSE did not improve for 5 consecutive epochs.

**Data Availability**
The layered intercalation compounds dataset and Materials Project alloy dataset used in this study are available from the corresponding author upon reasonable request. Pretrained GNN models (CGCNN, SchNet, DimeNet++) are available through the Open Catalyst Project.

**Code Availability**
Code for reproducing the analyses in this study is available from the corresponding author upon reasonable request.

**Acknowledgements**
We would like to acknowledge to Prof. Kiyou Shibata, Nagoya University for his helpful discussion. This study is studied by JST-Gtex.

**Author Contributions**
T.S. and T.M. designed the research. T.S. performed the calculations and analyzed the data. T.S. and T.M. wrote the manuscript.

**Competing Interests**
The authors declare no competing interests.

# Supplementary Materials of "Achieving Robust Extrapolation in Materials Property Prediction via Decoupled Transfer Learning"

Tasuku Sugiura and Teruyasu Mizoguchi

Institute of Industrial Science, The University of Tokyo, 153-8505 Tokyo, Japan

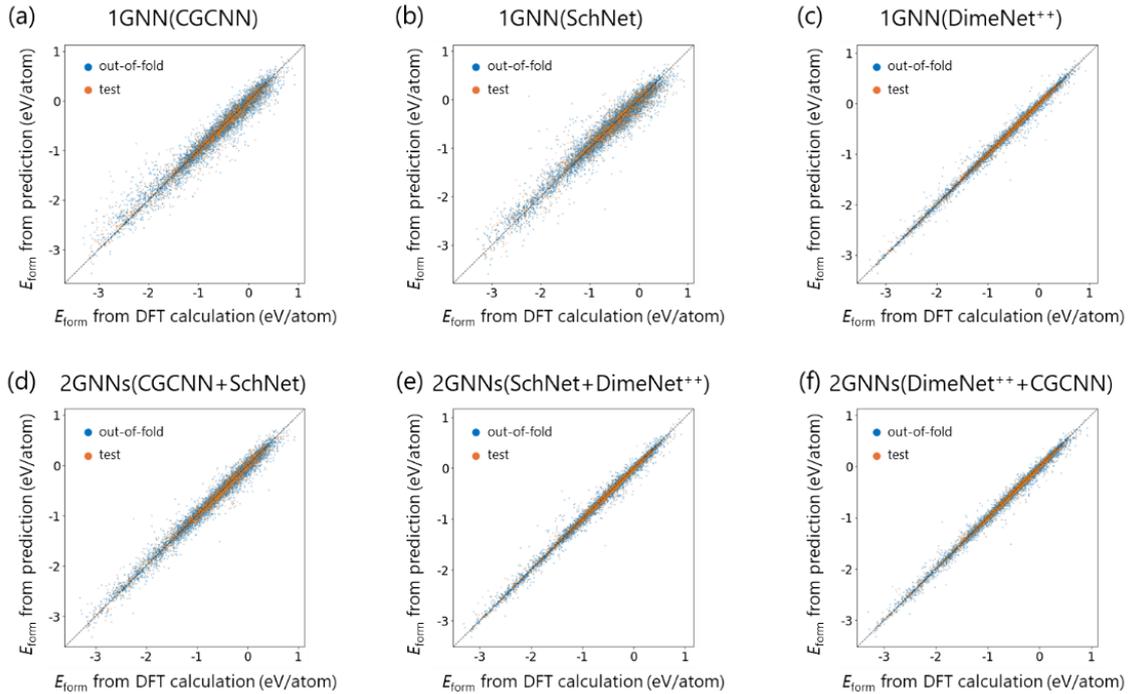

**Supplementary Fig. 1 | Prediction results for formation energy (eV/atom) under interpolation baseline for Layered Intercalation Compounds (LIC) dataset with single GNN+SVR and two GNNs+SVR decoupled approach.** (a-c) Single GNN+SVR configurations using CGCNN, SchNet, and DimeNet++, respectively. (d-f) Two GNNs+SVR combinations using CGCNN+SchNet, SchNet+DimeNet++, and CGCNN+DimeNet++, respectively. Blue and orange dots correspond to out-of-fold validation and test results, respectively. Results demonstrate that ensemble approaches with multiple GNN architectures provide complementary structural information, though single GNN configurations already achieve reasonable interpolation performance.

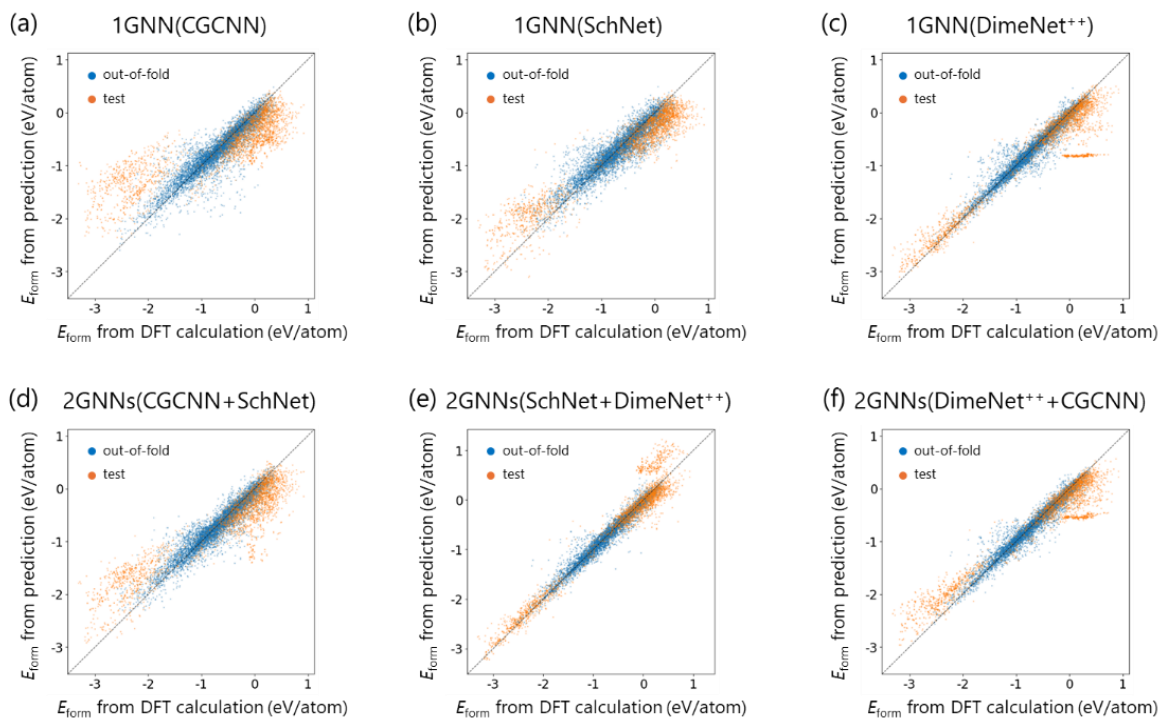

**Supplementary Fig. 2 | Prediction results for formation energy (eV/atom) under coupled extrapolation (structure and property) for Layered Intercalation Compounds (LIC) dataset with single GNN+SVR and two GNNs+SVR decoupled approach.** (a-c) Single GNN+SVR configurations using CGCNN, SchNet, and DimeNet++, respectively. (d-f) Two GNNs+SVR combinations using CGCNN+SchNet, SchNet+DimeNet++, and CGCNN+DimeNet++, respectively. Systematic comparison reveals that increasing the number of GNN architectures (from one to three) enhances extrapolation robustness by capturing complementary structural features. The combination of CGCNN (local coordination), SchNet (radial distributions), and DimeNet++ (angular information) provides the most comprehensive structural representation.

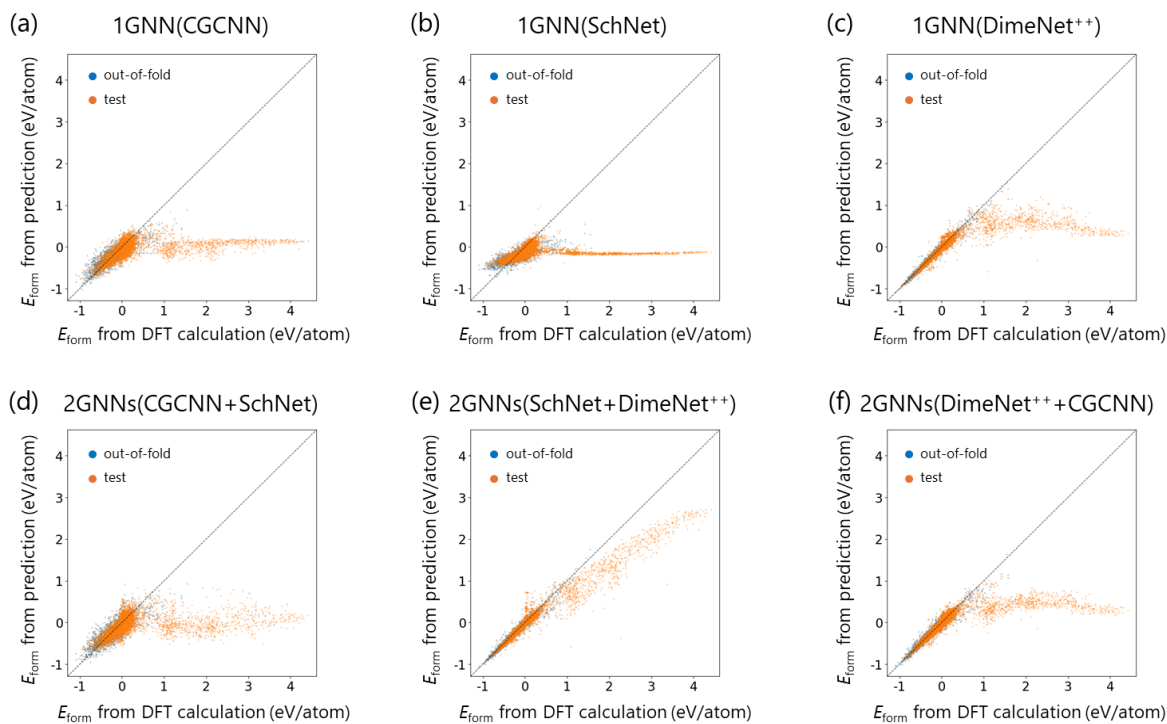

**Supplementary Fig. 3 | Prediction results for formation energy (eV/atom) for temporal extrapolation (MP18→MP21) Alloy of Interest (AoI) dataset with single GNN+SVR and two GNNs+SVR decoupled approach.** (a-c) Single GNN+SVR configurations using CGCNN, SchNet, and DimeNet++, respectively. (d-f) Two GNNs+SVR combinations using CGCNN+SchNet, SchNet+DimeNet++, and CGCNN+DimeNet++, respectively. Consistent with LIC dataset results, ensemble approaches combining multiple pretrained GNN architectures demonstrate superior extrapolation capability compared to single-architecture configurations, validating the generalizability of architectural complementarity across different materials classes and extrapolation scenarios.

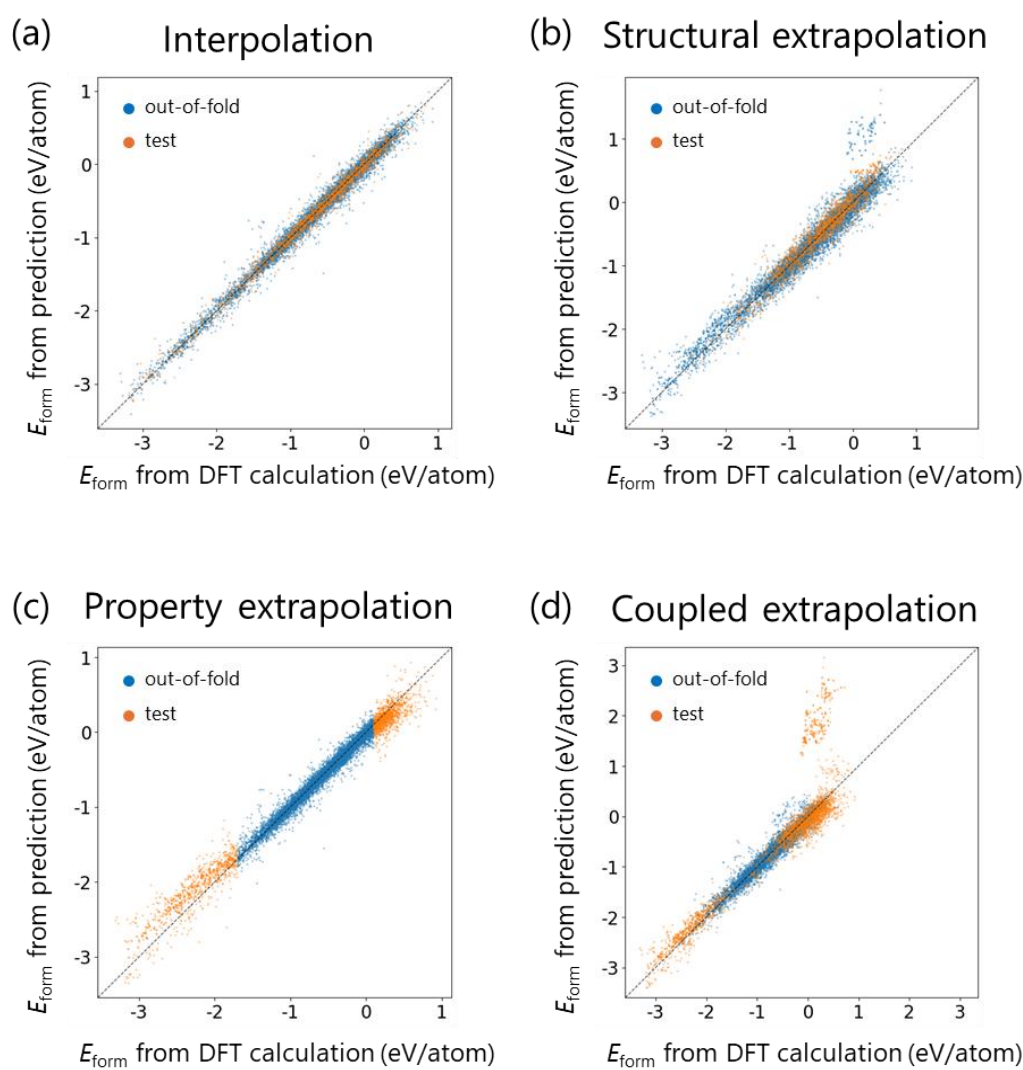

**Supplementary Fig. 4 | Prediction results using 3GNN+Ridge regression for the LIC dataset (formation energy per atom) across all four evaluation scenarios.** Ridge regression successfully outputs formation energy values outside the training range, demonstrating that simple linear regressors possess inherent extrapolation capability when provided with pretrained GNN features. Across (a) interpolation, (b) structural extrapolation, (c) property extrapolation, and (d) coupled extrapolation scenarios, Ridge achieves reasonable extrapolation performance, validating the decoupled transfer learning principle across different regression algorithms. However, comparison with 3GNN+SVR (main text, Figure 3) reveals that Ridge exhibits moderately higher errors, particularly in coupled extrapolation scenarios. This performance difference demonstrates the advantage of SVR with RBF kernel, which converts features to nonlinear space while maintaining extrapolation capability, providing enhanced robustness for materials with complex structural and electronic variations compared to purely linear regression.

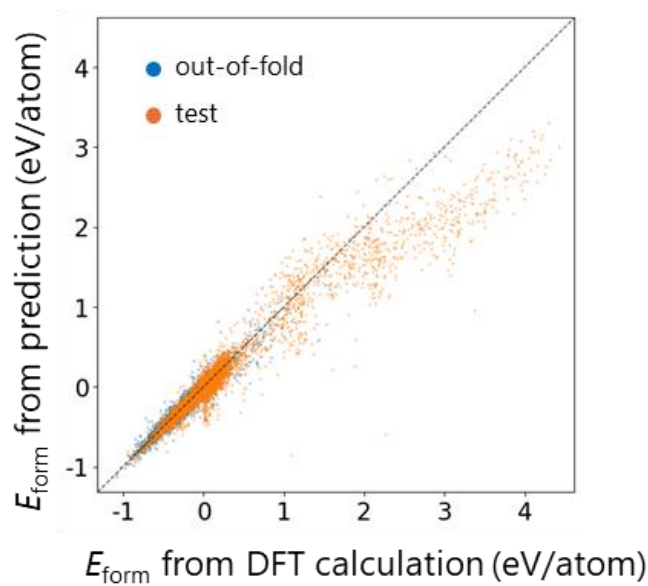

**Supplementary Fig. 5 | Prediction results using 3GNNs+Ridge method for the AoI temporal extrapolation dataset (formation energy per atom).** Similar to 3GNNs+SVR, 3GNNs+Ridge successfully extrapolates to formation energies outside the training range (-1.090 to 1.575 eV/atom), producing predictions extending to highly unstable structures (up to 4.416 eV/atom) in the MP21 test set. This confirms that simple linear regressors possess inherent extrapolation capability when provided with meaningful pretrained GNN features.

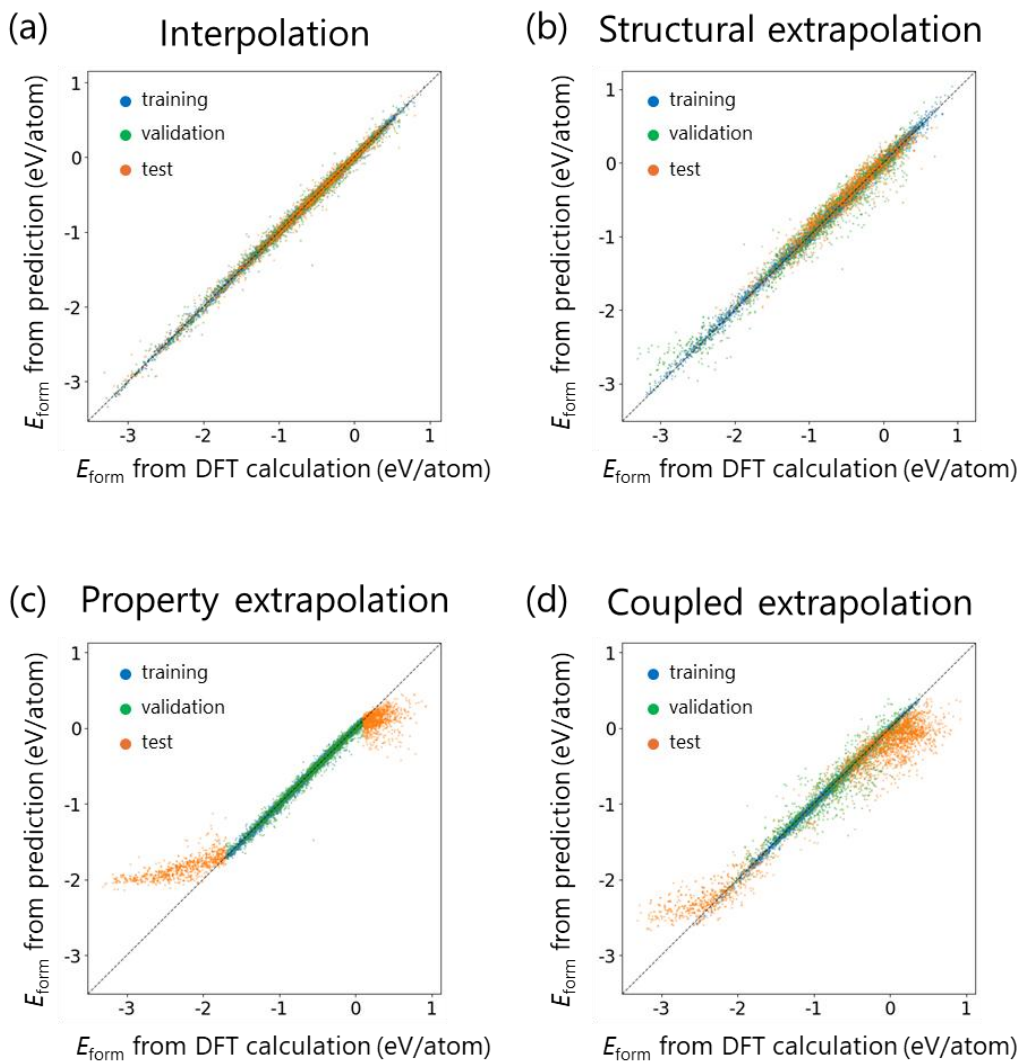

**Supplementary Fig. 6 | Prediction results using end-to-end fine-tuned CGCNN for the LIC dataset (formation energy per atom).** Baseline comparison across all four evaluation scenarios: (a) interpolation, (b) structural extrapolation, (c) property extrapolation, and (d) coupled extrapolation. End-to-end training achieves competitive interpolation performance but exhibits characteristic failure in property extrapolation scenarios (c, d), where predictions cluster near training range boundaries rather than extending to extreme formation energy values. This systematic failure, where the model cannot predict beyond the training property range, demonstrates the fundamental limitation of end-to-end approaches that couple feature learning with property prediction.

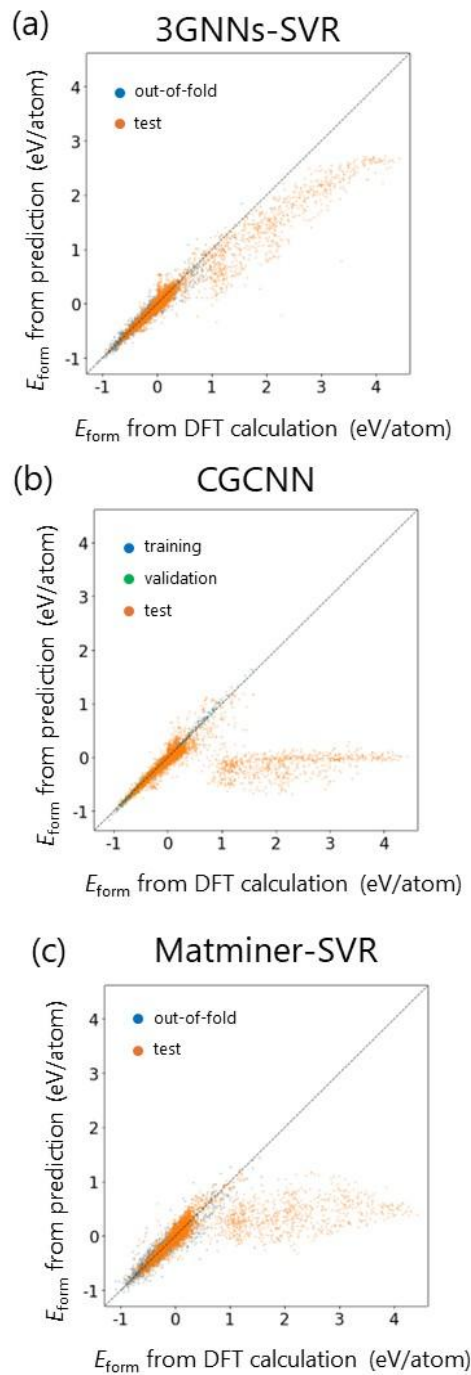

**Supplementary Fig. 7 | Comprehensive comparison of 3GNNs+SVR (present method), end-to-end CGCNN, and Matminer-SVR for the AoI dataset (formation energy per atom, temporal extrapolation).** Complete visualization showing out-of-fold validation, test predictions, and training/validation/test data distributions. Our decoupled transfer learning approach (3GNNs+SVR) achieves RMSE = 0.296 eV/atom, substantially outperforming both end-to-end CGCNN (RMSE = 0.876 eV/atom) and hand-crafted Matminer descriptors with SVR (RMSE = 0.702 eV/atom). Critically, while end-to-end CGCNN fails to produce outputs beyond ~2 eV/atom, our method maintains meaningful predictions across the full range to 4.416 eV/atom, validating the synergistic combination of pretrained GNN features and simple regression for extrapolation.

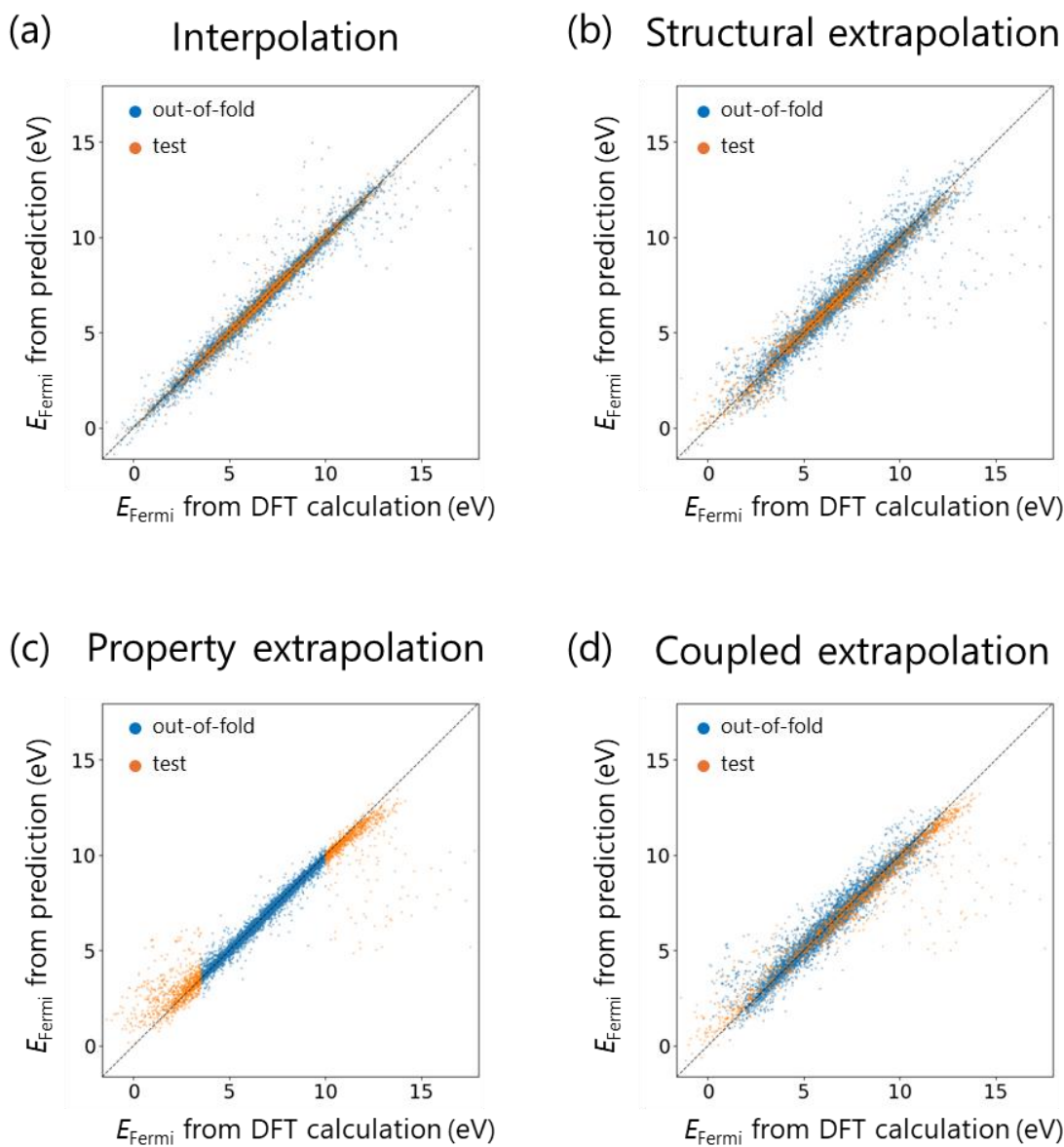

**Supplementary Fig. 8 | Extrapolation performance for Fermi energy prediction using the present method (3GNNs+SVR).** Identical decoupled transfer learning framework applied to DFT-calculated Fermi energies demonstrates comparable extrapolation capability across (a) interpolation, (b) structural extrapolation, (c) property extrapolation, and (d) coupled extrapolation scenarios. Successful application to this electronic property—fundamentally different from thermodynamic formation energy—demonstrates generalizability beyond formation energy to other single-structure DFT-computable properties. Test RMSE values: interpolation = 0.352 eV, structural extrapolation = 0.502 eV, property extrapolation = 1.270 eV, coupled extrapolation = 1.200 eV.

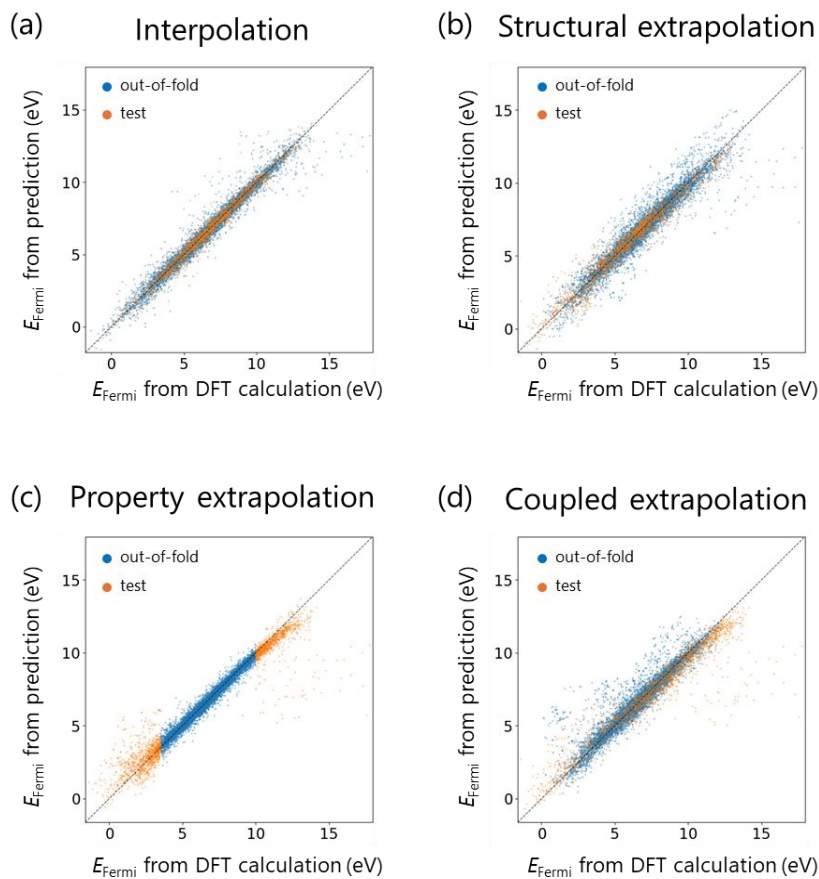

**Supplementary Fig. 9 | Extrapolation performance for Fermi energy prediction using Ridge regression (3GNNs+Ridge).** Alternative simple regressor applied to Fermi energy prediction across (a) interpolation, (b) structural extrapolation, (c) property extrapolation, and (d) coupled extrapolation scenarios. Ridge regression maintains extrapolation capability, confirming that simple linear models inherently extend predictions beyond training ranges when provided with pretrained GNN features. Performance is slightly inferior to SVR (Supplementary Fig. 8), particularly for coupled extrapolation, but substantially superior to end-to-end approaches, validating the decoupled transfer learning principle across different regression algorithms. Test RMSE values: interpolation = 0.436 eV, structural extrapolation = 0.575 eV, property extrapolation =1.229 eV, coupled extrapolation = 1.258eV.

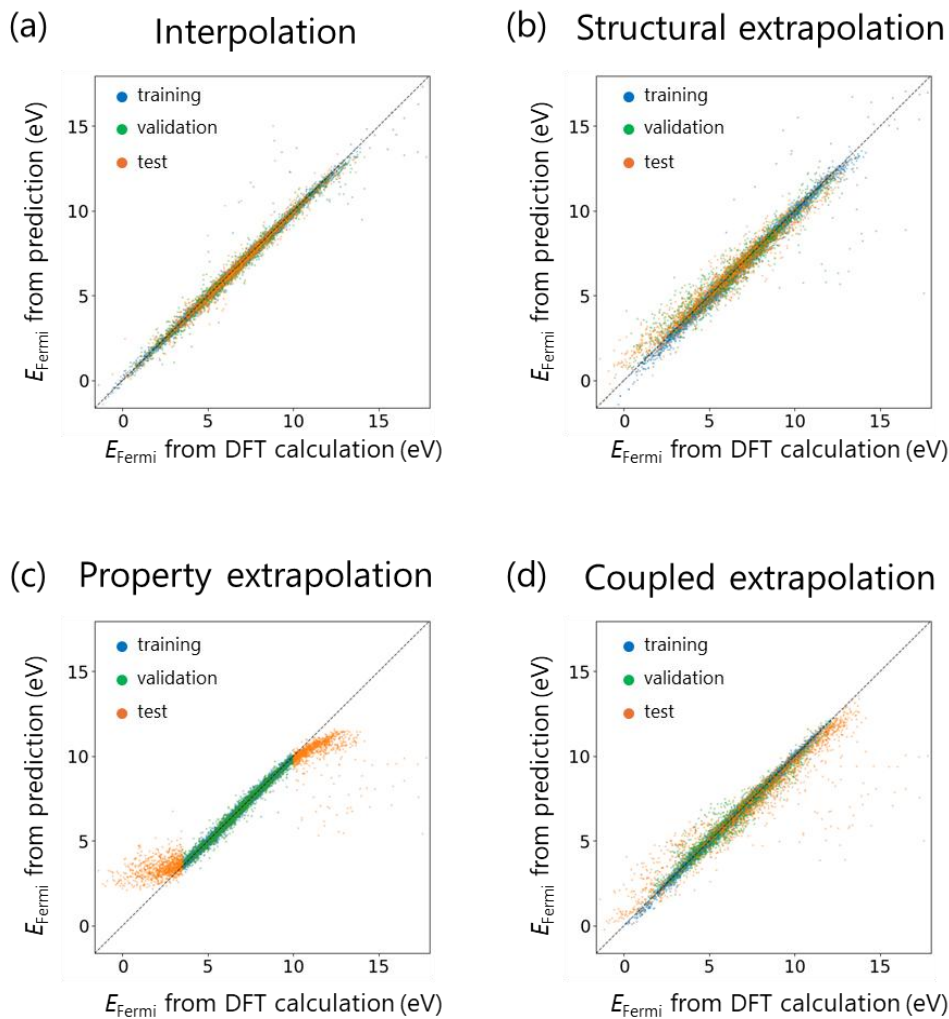

**Supplementary Fig. 10 | Extrapolation performance for Fermi energy prediction by end-to-end fine-tuned CGCNN.** Baseline comparison for Fermi energy across (a) interpolation, (b) structural extrapolation, (c) property extrapolation, and (d) coupled extrapolation scenarios. End-to-end CGCNN exhibits characteristic failure in property extrapolation (c, d), with predictions clustering within training range rather than extending to extreme Fermi energy values in the test set. This systematic collapse, consistent across both formation energy (main text, Supplementary Fig. 6) and Fermi energy predictions, confirms that coupled training fundamentally constrains output distributions regardless of target property type. Comparison with Supplementary Figs. 8-9 validates the superiority of decoupled transfer learning for extrapolative electronic property prediction.

**Supplementary Table 1 | Quantitative comparison of Fermi energy prediction performance on the layered intercalation compounds dataset.** Our decoupled transfer learning method (3GNNs+SVR) outperforms end-to-end GNN baseline (CGCNN) across all extrapolation scenarios. For structural extrapolation: MAE = 0.327 eV (present) vs. 0.455 eV (CGCNN); for property extrapolation: MAE = 0.641 eV vs. 1.043 eV; for coupled extrapolation: MAE = 0.612 eV vs. 0.755 eV. Consistent performance advantage across thermodynamic (formation energy, Table 1 in main text) and electronic (Fermi energy) properties establishes broad applicability of decoupled transfer learning to single-structure DFT-computable properties.

|  | Interpolation | | | Structural extrapolation | | | Property extrapolation | | | Coupled extrapolation | | |
| --- | --- | --- | --- | --- | --- | --- | --- | --- | --- | --- | --- | --- |
|  | MAE(eV) | RMSE (eV) | $R^2$ | MAE (eV) | RMSE (eV) | $R^2$ | MAE (eV) | RMSE (eV) | $R^2$ | MAE (eV) | RMSE (eV) | $R^2$ |
| **Present method (3GNNs+SVR)** | 0.188 | 0.352 | 0.981 | 0.327 | 0.502 | 0.967 | 0.641 | 1.269 | 0.923 | 0.612 | 1.200 | 0.885 |
| **End-to-end CGCNN** | 0.224 | 0.381 | 0.978 | 0.455 | 0.613 | 0.950 | 1.043 | 1.517 | 0.890 | 0.755 | 1.330 | 0.859 |